\documentclass[twocolumn]{webofc}
\usepackage[varg]{txfonts}   
\begin{document}
\title{Search for neutron-hidden neutron  interbrane transitions with MURMUR, a low-noise neutron passing-through-walls experiment}

\author{\firstname{Coraline} \lastname{Stasser}\inst{1}\fnsep\thanks{\email{coraline.stasser@unamur.be}} \and
        \firstname{Michaël} \lastname{Sarrazin}\inst{1,2}\fnsep\thanks{\email{michael.sarrazin@unamur.be}} \and
        \firstname{Guy} \lastname{Terwagne}\inst{1}\fnsep\thanks{\email{guy.terwagne@unamur.be}}
}

\institute{The MURMUR collaboration, Laboratory of Analysis by Nuclear Reaction, Department of physics, University of Namur, 61 rue de Bruxelles, B-5000 Namur, Belgium \and Institut UTINAM, CNRS/INSU, UMR 6213, Université Bourgogne-Franche-Comté, 16 route de Gray, F-25030 Besançon Cedex, France}

\abstract{Multi-braneworld universe is at the heart of many scenarios of physics beyond the Standard Model and the cosmological model $\Lambda CDM$. It is thus an important concern to constrain these scenarios which also allow for experiments below the GeV scale. MURMUR is a new neutron passing-through-walls experiment designed to constrain neutron-hidden neutron transitions in the context of braneworlds scenarios. As our visible universe could be a 3-brane embedded in a multidimensional bulk, adjacent hidden 3-branes are often expected. Their existence can be constrained as matter exchange between braneworlds must occur with a swapping probability $p$. A neutron $n$ can convert into a hidden neutron $n’$ when scattered by a nucleus with cross section $\sigma (n \to n')$ $\propto$ $\sigma_E (n \to n)\  p$, where $\sigma_E$ is the usual elastic cross-section. Hidden neutrons could therefore be generated in the moderator medium of a nuclear reactor, where high-flux neutrons undergo many elastic collisions. All the theoretical and technological keys of this experiment soon to be installed at the nuclear research reactor BR2 (SCK.CEN, Mol, Belgium) are introduced.

}

\maketitle

\section{Introduction}
\label{intro}
The universe may possess more than the three spatial dimensions than we can perceive. This idea has prompted a great interest in the scientific community in the last decades. Following this new hypothesis, a lot of theories and models were established. All of them have one commonality: finding the nature of the new physics brings to light from the limits of the Standard Model of particles.  This is how the concept of braneworld emerges as a 3D sheet of space embedded in a universe called the bulk with $N>4$ dimensions. The original concept of braneworld appears in string theory \cite{ref10} and reaches its culmination with the Horava-Witten approach \cite{ref19}. However, braneworld physics does not remain limited exclusively to string theory. Indeed, some theories often consider a brane as a topological default of the bulk endowed with one (or two) large extra dimension, and explain how particles are located on these defects \cite{ref8,ref11}. Along this line of thought, many models emerge trying to solve the hierarchy problem \cite{ref13} or to offer an alternative to cosmic inflation \cite{ref12} or an explanation of dark energy and dark matter \cite{ref16,ref17,ref18}. In some models, one considers a two-brane universe in which the visible and the hidden branes both sustain a copy of the Standard Model of particles. Such a scenario may lead to a violation of the Lorentz invariance on branes when a specific metric is employed, like a Chung-Freese metric \cite{ref20}. This approach offers a non-inflationary solution to the cosmological horizon problem \cite{ref17,ref18}. It is thus a major concern to experimentally constrain this kind of scenario.


\noindent In 2010, one of us (M. Sarrazin) showed that matter exchanges should occur at low energy between braneworlds in the bulk, as long as they are close enough to each other \cite{ref1}. Thus, neutron swapping is allowed between our visible brane and a hidden one.  The ability for a neutron to leap from our visible brane to reach a hidden one and reciprocally is given by a probability $p$. Such an effect is probed in some experiments involving search for passing-through-walls neutrons in the vicinity of a nuclear reactor \cite{ref2,ref6} or search for unusual leaks when dealing with ultracold neutron storage \cite{ref21}. The aim of this paper is to describe a new neutron passing-through-walls experiment in preparation at the University of Namur and soon to be installed near the BR2 reactor of the SCK.CEN (Mol, Belgium). This experiment, called MURMUR,  possesses the particularity to use lead as an antenna to regenerate hidden neutrons into visible ones.

\section{Two-brane Universe modeled by a noncommutative two-sheeted $\mathbf{M_{4}         \bigotimes Z_{2}}$ spacetime}
\label{sec-1}
\begin{figure}[h]
\centering
\includegraphics[width=8cm,clip]{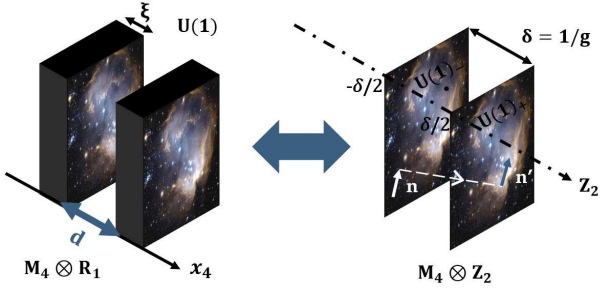}
\caption{Equivalence between a two-braneworld system and a two-sheeted   spacetime. This framework makes possible to demonstrate the matter exchange between branes. $d$ is the interbrane distance in the bulk, and $\xi$ the brane thickness.}
\label{fig-1}       
\end{figure}

This model \cite{ref1} allows the fermion dynamics description at low energy for a two-brane universe. At low energies, any two-braneworld system is equivalent to a two-sheeted noncommutative spacetime $\mathbf{M_{4}\bigotimes Z_{2}}$. Sheets are then separated by an effective distance $\delta=1/g$, where $g$ is the interbrane coupling constant between fermion fields. The bulk gauge field $U(1)$ is substituted by $U_+(1)$ and $U_-(1)$ associated with gauge fields confined in 4 dimensions respectively on the brane $+$ and $-$ (see figure \ref{fig-1}).

\noindent This description at low energy is valid regardless of the nature of the branes and the properties of the bulk, being then a general framework. This description leads to a $\mathbf{M_{4}\bigotimes Z_{2}}$ Dirac equation. The non-relativistic limit of this one makes possible to bring to light matter swapping between the two branes, and particularly neutrons swapping given by the following probability \cite{ref1}

\begin{equation}
p=\frac{2\Omega^2}{\eta^2},  \label{p}
\end{equation}
with 
\begin{equation}
\eta=\frac{V_+-V_-}{\hbar},  \label{eta}
\end{equation}
and 
\begin{equation}
\Omega=\frac{g \, \mu_n \, A_{amb}}{\hbar},  \label{omega}
\end{equation}

\noindent where $V_+-V_-$ and $A_{amb}$ respectively are the neutron gravitational energy and the magnetic vector potential differences between the two branes and $\mu_n$ the neutron magnetic moment. It is possible to constrain $p$ thanks to neutron passing-through-walls experiments. The last one, realized at the ILL (Grenoble, France) led to \cite{ref2}
\begin{equation}
p<4.6\times10^{-10} \ \text{at 95\% CL.}  \label{plimit}
\end{equation}

\begin{figure}[h]
\centering
\includegraphics[width=8.5cm,clip]{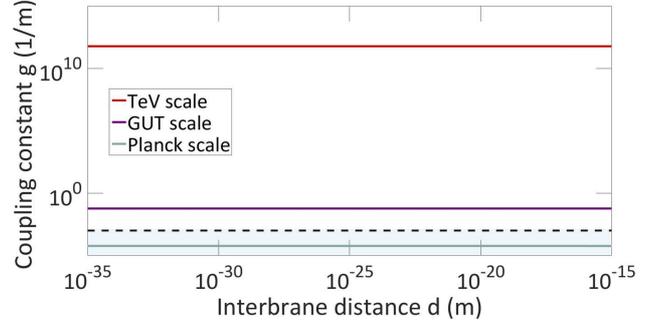}
\caption{Coupling constant $g$ limits against interbrane distance $d$ in a 5D bulk for various energy scales $2 \ M_B \ (R^{-3/2}+R^{3/2})^{-1}$ of the branes including GUT and Planck energy scales  \cite{ref4}. For such distances $d$, the influence of the exponential function disappears from the expression (\ref{CF g}). The dashed black line represents the latest constraint on $g$ \cite{ref2}. The blue region is thus expected to be likely reachable by neutron passing-through-walls experiment like MURMUR. Only the Planck scale is reachable in such a scenario. This result is promising because it shows that Planck scale is accessible with neutron passing-through-walls experiment like MURMUR. 
}
\label{fig-2}       
\end{figure}

The coupling constant $g$ must depend on the interbrane distance $d$, the brane thickness $\xi$ and on the bulk metric. It is possible to derive the explicit expression of $g$ thanks to a string-inspired low energy classical field model \cite{ref4}. This approach allows to describe the quark dynamics at low energy in a two-brane Universe and allows for the reconstruction of the coupling constant $g$ for a neutron thanks to the quark constituent model \cite{ref5}. The explicit expression of the coupling constant  $g$ between two braneworlds endowed with their own copy of the standard model in a $\mathbf{M_{4} \bigotimes R_{1}}$ bulk with a Chung-Freese \cite{ref3} metric is given by \cite{ref4}

\begin{equation}
g=\frac{1}{2} \frac{m^2}{M_B} e^{-md}\ (R^{-3/2}+R^{3/2}),  \label{CF g}
\end{equation}

\noindent where $m$ is the mass of a constituent quark (340 MeV) \cite{ref5}, $M_B$ is the energy scale of the branes, such that $M_B^{-1}$ corresponds to the brane thickness, $R$ the ratio of the scale factors of each brane and $d$ the interbrane distance. Figure \ref{fig-2} shows the limits of the coupling constant $g$ against interbrane distance $d$ in a 5D bulk for various energy scales of the branes including GUT and Planck scales \cite{ref4}. 

\section{Passing-through-walls neutrons}
\begin{figure}[h]
\centering
\includegraphics[width=9cm,clip]{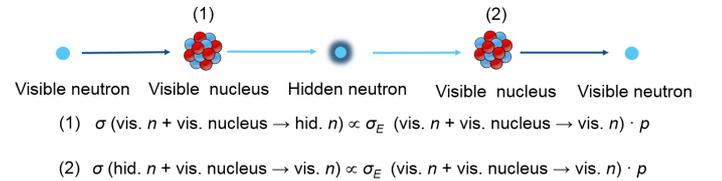}
\caption{Passing-through-walls neutrons at a nuclear level.}
\label{fig-3}       
\end{figure}

Neutron disappearance/reappearance toward/from a hidden brane can be tested with high-precision experiments \cite{ref6}. Neutron/nucleus elastic collisions induce hidden neutrons. This process is reversible (see figure \ref{fig-3}).

\begin{figure}[h]
\centering
\includegraphics[width=8cm,clip]{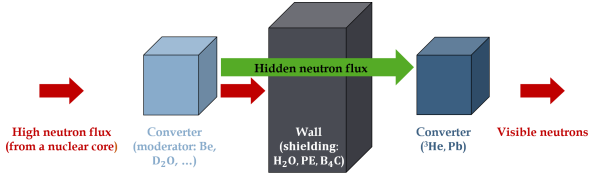}
\caption{Neutron passing-through-walls experimental setup.}
\label{fig-4}       
\end{figure}

\noindent As summarized in figure \ref{fig-4}, a neutron passing-through-walls experiment consists into a high neutron flux, a first converter which transforms visible neutrons into hidden neutrons, a shielding that plays the role of a wall for visible neutrons while letting hidden neutrons free to pass through and a second converter able to regenerate hidden neutrons into visible ones to be finally detected thanks to a neutron counter. This setup gives rise to two source terms $S_{-}$ and $S_{+}$, respectively the hidden and the regenerated neutron source terms, both given by the expression \cite{ref6}: 

\begin{equation}
S_{\mp}=\frac{1}{2}\ p \ \Sigma_{converter} \ \Phi_{\pm},  \label{Source}
\end{equation}

\noindent where $p$ is the neutron swapping probability, $\Sigma_{converter}$ the macroscopic elastic cross section of the corresponding converter (see figure \ref{fig-4}), $\Phi_+$ the incident neutron flux coming from the nuclear core and $\Phi_-$ the hidden neutron flux generated by the first converter. 

\section{MURMUR experiment}
The MURMUR project is a new neutron passing-through-walls experiment soon to be installed near the BR2 nuclear reactor of the SCK.CEN (Mol, Belgium). The aim of this experiment is to constrain some braneworld cosmological scenarios \cite{ref3} and fix a new stronger limit on the neutron swapping probability $p$.

\subsection{Design of the MURMUR experiment}
\begin{figure}[]
\centering
\includegraphics[width=8cm,clip]{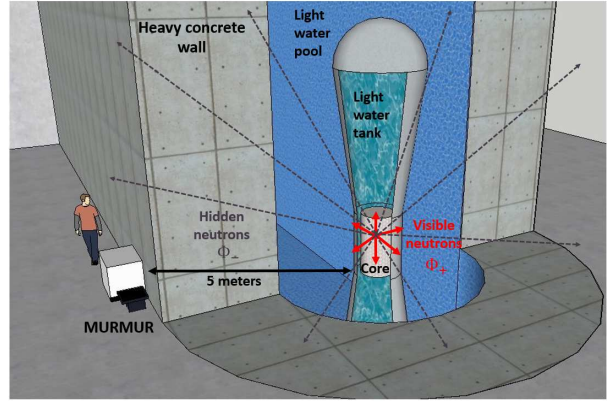}
\caption{Geometry of the MURMUR experiment near the BR2 nuclear reactor of the SCK.CEN (Mol, Belgium).}
\label{fig-5}       
\end{figure}

\begin{figure}[]
\centering
\includegraphics[width=8cm,clip]{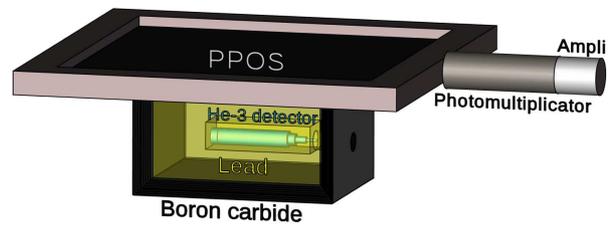}
\caption{Full detector device of the MURMUR experiment, constituted by a $36$ $cm^3$ He-3 proportional counter embedded in a lead matrix of $24 \times 18 \times 12$ $cm^3$ and a $4 \pi$ boron carbide shielding of $35$ $mm$ of thickness. A plastic plan organic scintillator (PPOS) EJ200 of $50 \times 50$ $cm^2$ monitors the cosmic muon flux and simultaneously veto the data acquisition.}
\label{fig-6}       
\end{figure}

MURMUR uses a lead antenna as regenerator material. Indeed, lead appears to be an ideal material thanks to its high elastic cross section ($0.367\ cm^{-1}$) and its low neutron absorption. Thus, hidden neutrons can be regenerated into visible neutrons in a large volume of lead without being absorbed. Regenerated visible neutrons are free to propagate in lead up to an He-3 detector put in the center of the lead. The moderator of the nuclear core, a beryllium matrix, acts as visible neutrons/hidden neutrons converter. A boron carbide ($B_4C$) $4\pi$-shield absorbs visible neutrons from background and isolates the lead antenna and the  He-3 detector  from the external environment. A plane plastic organic scintillator (PPOS) EJ200 is placed on the top of the shielding in order to veto the data acquisition. Indeed, cosmic muons induce spallation neutrons inside lead, which must be cancelled. Lead also minimizes neutron thermalization thus avoiding noise from fast neutrons related to the cosmic rays background. Figures \ref{fig-5} and \ref{fig-6} show respectively the geometry of the MURMUR experiment near the BR2 reactor at Mol and the disposition of the different elements which constitute MURMUR. 

\subsection{Data acquisition chain}
\begin{figure}[h]
\centering
\includegraphics[width=8cm,clip]{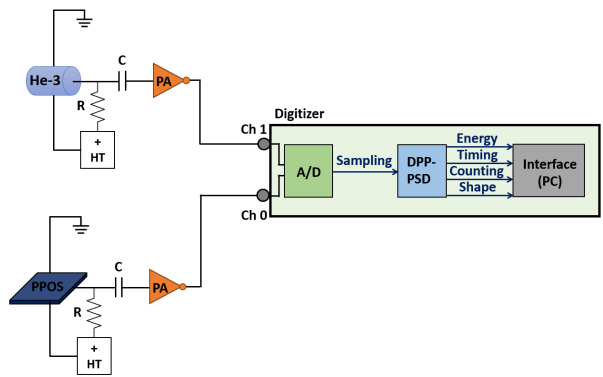}
\caption{Data acquisition chain of the MURMUR experiment.}
\label{fig-7}       
\end{figure}

The scheme of the data acquisition of the MURMUR experiment is presented in figure \ref{fig-7}. The acquisition control software is a modified DPP-PSD code of the CAEN firmware. Some additional functionalities were added to make it suitable for the MURMUR experiment.  An anti-coincidence mode is enabled between the two detectors in order to veto the He-3 detector. Each time an event is measured in the He-3 detector in a time window of $3$ $\mu s$ after one event was measured in the scintillator, it is automatically suppressed by the software. A veto is essential to overcome the muon-induced neutrons in the lead, main source of noise of the MURMUR experiment.  

\subsection{Background noise hunting}

The challenge of the MURMUR experiment is to reach a very low level of noise ($10^{-5}\ Hz$). For that, some precautions have to be taken. First, the He-3 proportional counter and the lead are embedded in a   $4 \pi$ boron carbide shield in order to absorb the visible neutrons background and isolate the lead and He-3 detector from the environment. Then, a plane plastic organic scintillator is employed for vetoing the acquisition in order to reject fast neutrons induced by cosmic muon spallation in lead. It makes possible to reduce the noise of MURMUR by 96.7\%. Pulse shape discrimination is also used to discriminate $\gamma$ rays in the aim to improve signal/noise ratio. Finally, the most important one is the realization of ON/OFF reactor measurements with and without the antenna in order to exclude  cosmic rays and internal noise and to identify the efficiency of the lead. Thus, if a signal is measured by MURMUR, there are 3 possible origins: fast neutrons thermalized by the antenna, neutrons induced by neutrinos capture in the antenna and hidden neutron detection.

\subsection{Neutron fluxes computations}
In order to extract the swapping probability $p$ from the neutron event rate measured in the He-3 detector, it is necessary to compute the neutron fluxes around the reactor and in the MURMUR device as a function of $p$. The following expressions give the hidden and regenerated neutron fluxes, respectively $\Phi_-$, $\Phi_R$, from equation \ref{Source}  \cite{ref6}:

\begin{equation}
\Phi _{-}(\mathbf{r})=\frac p{8\pi }\int_{Tank}\frac 1{\left| \mathbf{r-r}%
^{\prime }\right| ^2}\Sigma _{Be}(\mathbf{r}^{\prime }) \, \Phi _{+}(\mathbf{r}%
^{\prime }) \, d^3r^{\prime },  \label{Phi+}
\end{equation}

\begin{equation}
\Phi _{R}(\mathbf{r})=\frac p{8\pi }\int_{Antenna}\frac 1{\left| \mathbf{r-r}%
^{\prime }\right| ^2}\Sigma _{Pb}(\mathbf{r}^{\prime }) \, \Phi _{-}(\mathbf{r}%
^{\prime }) \, d^3r^{\prime },  \label{PhiR}
\end{equation}

\noindent where $\Phi_+$ is the neutron flux operated inside the core, $r$ is the distance from the nuclear core, $\Sigma_{Be}$ and  $\Sigma_{Pb}$ are the macroscopic elastic cross-sections of the beryllium and the lead respectively, $p$ is the neutron swapping probability. Looking at equation \ref{PhiR}, we can deduce that the expected phenomenon is proportional to $p^2$. While $\Phi_+$  can be calculated with ab initio numerical computations, $\Phi_R$ can be deduced from the neutron rate measured in the detector. 


\subsection{Improvements and limitations of the MURMUR experiment}

\noindent The MURMUR experiment is an improved version of the neutron passing-through-wall experiment realized at the ILL in 2015 \cite{ref6}, which makes possible to put a constraint on the neutron swapping probability $p$ (see equation \ref{plimit}). The first and most important improvement is the addition of an antenna made of lead in order to regenerate hidden neutrons into visible ones. In the first experiment, it was only the He-3 of the neutron detector which ensured the regeneration of the hidden neutrons, detecting them at the same time. The efficiency of a neutron passing-through-wall experiment is proportional to the volume of the regenerating material. MURMUR has an effective volume of regenerating material 122 times bigger than the first experiment at the ILL. The second improvement is the choice of the BR2 reactor of the SCK.CEN at Mol which gives a less noisy environment. Indeed, while there are neutron guides out of the reactor at the ILL, that is not the case at the BR2. Furthermore, the emplacement near the BR2 makes possible to realize two years of measurements as well as reactor ON and reactor OFF acquisitions. It is then a major enhancement in comparison with the first experiment at the ILL which only ensured $44$ hours of reactor ON measurements \cite{ref2}. The last improvement is the addition of a scintillator and the possibility to carry out Pulse Shape Discrimination in order to reduce the background noise. Thanks to all these enhancements, one expects to reach an improvement of the neutron swapping probability constraint of at least one order of magnitude. Future neutron passing-through-wall experiments, with more lead and bigger He-3 detector for instance, could lead to lower value of this constraint.


\section{Conclusion}

MURMUR is a new high sensitivity neutron passing-through-walls experiment which the aim to fix a new stronger limit on the neutron swapping probability $p$ in order to constrain some braneworld scenarios. The data acquisition of MURMUR will begin on October 2018 near the BR2 reactor of the SCK.CEN at Mol in Belgium. This kind of experiment is promising to constrain braneworld scenarios and, thus, the existence of extra dimensions close to the Planck energy scale \cite{ref4}. 
\medbreak

\section*{Acknowledgements}

This presentation was supported by the MURMUR Collaboration (http://www.murmur-experiment.eu) and the University of Namur. The authors gratefully acknowledge the members of the collaboration for useful preliminary discussions about the project.

%
%
%

\end{document}